\documentclass[journal,onecolumn,draftcls,12pt]{IEEEtran}
\usepackage[T1]{fontenc}

\makeatletter
\def\endthebibliography{%
	\def\@noitemerr{\@latex@warning{Empty `thebibliography' environment}}%
	\endlist
}
\makeatother

\IEEEoverridecommandlockouts
\usepackage{cite}
\usepackage{amsmath,amssymb,amsfonts}
\usepackage{algorithm}
\usepackage{algorithmic}
\usepackage{etoolbox}  
\usepackage{bbm}
\usepackage{amsmath}
\usepackage{tabularx}
\usepackage{amssymb}
\usepackage{threeparttable}
\usepackage{booktabs}
\usepackage{amsthm}
\usepackage{subcaption}
\usepackage{bm}
\makeatletter
\patchcmd{\algorithmic}{\addtolength{\ALC@tlm}{\leftmargin} }{\addtolength{\ALC@tlm}{\leftmargin}}{}{}
\makeatother

\makeatletter
\newcommand\fs@betterruled{%
	\def\@fs@cfont{\bfseries}\let\@fs@capt\floatc@ruled
	\def\@fs@pre{\vspace*{5pt}\hrule height.8pt depth0pt \kern2pt}%
	\def\@fs@post{\kern2pt\hrule\relax}%
	\def\@fs@mid{\kern2pt\hrule\kern2pt}%
	\let\@fs@iftopcapt\iftrue}
\floatstyle{betterruled}
\restylefloat{algorithm}
\makeatother

\usepackage{tikz}
    \usetikzlibrary{shapes.arrows}

\usepackage{pgfplots}
\usepackage{adjustbox}

\usepackage{gincltex}
\usepgfplotslibrary{fillbetween}
\usetikzlibrary{plotmarks}
\usetikzlibrary{patterns}

\usepackage{filecontents}
\usepackage{graphicx}
\usepackage{textcomp}
\usepackage{xcolor}
\usepackage{authblk}
\def\BibTeX{{\rm B\kern-.05em{\sc i\kern-.025em b}\kern-.08em
		T\kern-.1667em\lower.7ex\hbox{E}\kern-.125emX}}

\pgfplotsset{compat=1.14}
		
\usepackage{glossaries}

\newacronym{opf}{OPF}{Oldest Packet First}
\newacronym{isl}{ISL}{Inter-Satellite Link}
\newacronym{jfi}{JFI}{Jain Fairness Index}
\newacronym{aoi}{AoI}{Age of Information}
\newacronym{paoi}{PAoI}{Peak Age of Information}
\newacronym{pdf}{PDF}{Probability Density Function}
\newacronym{mpr}{MPR}{Multi-Packet Reception}
\newacronym{cdf}{CDF}{Cumulative Density Function}
\newacronym{fcfs}{FCFS}{First Come First Serve}
\newacronym{lcfs}{LCFS}{Last Come First Serve}
\newacronym{haf}{HAF}{Highest Age First}
\newacronym{iot}{IoT}{Internet of Things}
\newacronym{mdp}{MDP}{Markov Decision Process}
\newacronym{ais}{AIS}{Authentication Identification System}
\newacronym{vdes}{VDES}{VHF Data Exchange System}
\newacronym{leo}{LEO}{Low Earth Orbit}
\newacronym{geo}{GEO}{Geo-synchronous Equatorial Orbit}
\newacronym{meo}{MEO}{Medium Earth Orbit}
\newacronym{e2e}{E2E}{end-to-end}
\newacronym{ml}{ML}{Machine Learning}
\newacronym{unb}{UNB}{Ultra-Narrowband}
\newacronym{adsb}{ADS-B}{Automatic Dependent Surveillance - Broadcast}
\newacronym{dl}{DL}{downlink}
\newacronym{ul}{UL}{uplink}

\definecolor{violet}{rgb}{0.6,0,0.6}%
\definecolor{orange_D}{rgb}{1,0.3,0}%
\definecolor{cyan}{rgb}{0,0.67,0.64}%
\definecolor{red}{rgb}{0.9,0,0}%
\definecolor{green}{rgb}{0,0.8,0}%
\definecolor{yellow}{rgb}{1,0.8,0}

\def \fwidth{0.7\columnwidth}
\def \fheight {0.32\columnwidth}

\begin{document}

\title{Information Freshness of Updates Sent over LEO Satellite Multi-Hop Networks}

\author{Federico Chiariotti,~\IEEEmembership{Member,~IEEE,}
        Olga Vikhrova,
        Beatriz Soret,~\IEEEmembership{Member,~IEEE,}
        and~Petar~Popovski,~\IEEEmembership{Fellow,~IEEE}
\thanks{F. Chiariotti (corresponding author, email: fchi@es.aau.dk), B. Soret, and P. Popovski are with the Department of Electronic Systems, Aalborg University, 9100 Aalborg, Denmark. O. Vikhrova is with DIIES Department, University Mediterranea of Reggio Calabria, 89100 Reggio Calabria, Italy.}}

\maketitle

\begin{abstract}
\gls{leo} satellite constellations are bringing the \gls{iot} to the space arena, also known as non-terrestrial networks. Several \gls{iot} satellite applications for tracking ships and cargo can be seen as exemplary cases of intermittent transmission of updates whose main performance parameter is the information freshness. This paper analyzes the \gls{aoi} of a satellite network with multiple sources and destinations that are very distant and therefore require several consecutive multi-hop transmissions. A packet erasure channel and different queueing policies are considered. We provide closed-form bounds and tight approximations of the average \gls{aoi}, as well as an upper bound of the \gls{paoi} distribution as a worst-case metric for the system design. The performance evaluation reveals complex trade-offs among age, load, and packet losses. The optimal operational point is found when the combination of arrival rates and packet losses is such that the system load can ensure fresh information at the receiver; nevertheless, achieving this is highly dependent on the mesh topology. Moreover, the potential of an age-aware scheduling strategy is investigated and the fairness among users discussed. The results show the need to identify the bottleneck nodes for the age, as improving the rate and reliability of those critical links will highly impact on the overall performance. The model is general enough to represent other multi-hop mesh networks.
\end{abstract}


\IEEEpeerreviewmaketitle
\glsresetall

\section{Introduction}

Satellite communications are characterized by the inherent delay due to the large physical distances. Such propagation delay is highly reduced when using \gls{leo} satellites with altitudes between 500 and 2000 km, and propagation delays in the order of milliseconds. Unlike geostationary orbits, \gls{leo} satellites move fast with respect to the Earth's surface and have a small ground coverage: only 0.45 \% of the Earth's surface for a LEO satellite deployed at 600 km and with an elevation angle of 30 degrees. To ensure that any ground terminal is always covered by, at least, one satellite, a flying formation of many satellites is required, usually organized in a \emph{constellation} with coordinated ground coverage~\cite{walker1971circular,Soret2020}.

Latency-sensitive information might suffer from long delays even with low orbits, because several inter-connected satellites are required to connect two distant points on the Earth's surface. The result is a multi-hop network where intermediate nodes (satellites) along the path receive and forward packets via wireless links. The introduction of multi-hop connectivity has the drawback of additional latency \cite{Khanna2018,Yang2018}, as the total latency is a combination of processing delay, queueing delay, transmission time and propagation delay at each hop. 

Besides latency, a related quantity of interest that has recently attracted significant information is the \gls{aoi} metric \cite{kaul2012real}. \gls{aoi} is defined as the time elapsed since the last received message containing update information was generated and it can be interpreted as a representative of the freshness of the sensory information at the receiver. Many \gls{iot} applications that rely on satellites involve tracking of e.g. ships or cargo, such that an \gls{iot} transmission in this setting is often a real-time status update. Hence, satellite \gls{iot} communication entails some of the exemplary cases where information freshness and \gls{aoi} are of primary importance, rather than the conventional latency or packet delay. These examples include the recently-introduced  \gls{vdes} \cite{ITU2371}\cite{Lazaro2019} for maritime communications and its predecessor \gls{ais}. The \gls{paoi} \cite{Huang2015} is a byproduct of the age process that quantifies the worst case. 

\begin{figure}[t]
	\centering
	\includegraphics[width=4.5in]{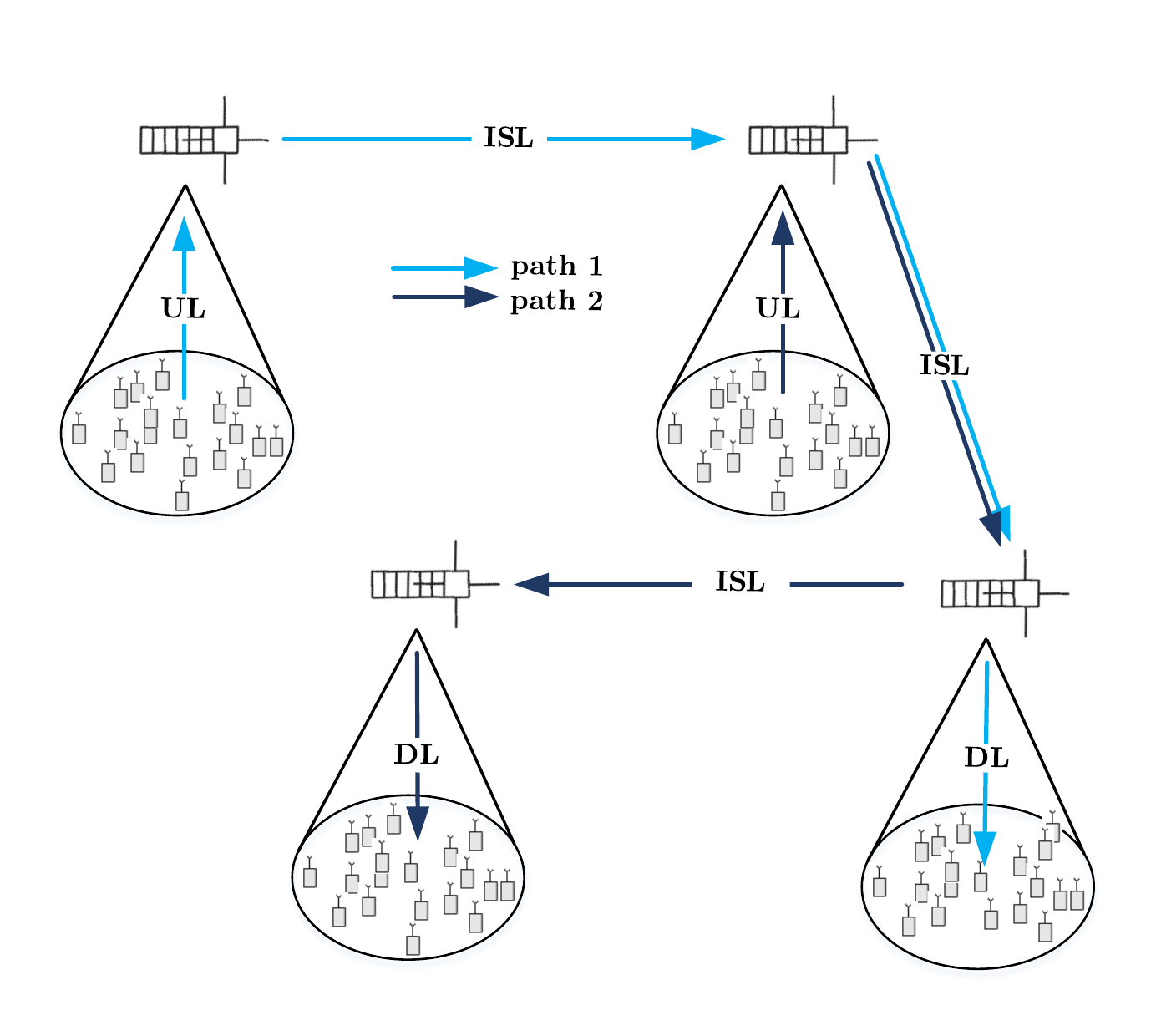}
	\caption{Example of a multi-hop relaying satellite network with multiple sources and destinations. There are a total of four nodes and two exemplary path routes indicated in light blue and dark blue. }
	\label{fig_scenario}
\end{figure} 

In this paper, we address a general buffer-aided multi-hop network with multiple sources and destinations. 
We focus on applications where a number of mutually independent traffic sources need to report sensory updates to a number of control stations in a timely fashion, and do so through a chain of \gls{leo} satellites. Since the end-points are too far to be linked by a single connection, several satellites connected by \glspl{isl} are required to relay the information to the final point.

As Fig.~\ref{fig_scenario} shows, all the satellites work as a relay and, at the same time, receive uplink status-updates from their coverage area and are the final destination for some of the status-updates. At each satellite, the \gls{isl} is used to forward both the ground information from satellite and the packets from neighboring satellites. Different sources can have partly overlapping paths, using the same relays for part of their connection to their ground destinations, which can be ground stations, gateways, or ground devices with direct connectivity to the satellite network. The \gls{isl} links, the uplink from the source to the first satellite, and the downlink from the last satellite to the ground destination can have different capacities and packet loss rates, due to the different technologies, propagation environments, and power budgets~\cite{Soret2020,popescu2017power}.

These multi-hop networks are difficult to study theoretically due to the complex interactions between subsequent queueing systems, and the literature on the subject is limited. 
In this queueing network, the server represents the wireless links whose rates are different, and each node works as a relay for the previous node and receives data packets from its directly connected traffic source. These two main traits of the model greatly complicate the queueing system. Furthermore, the presence of errors in the transmission of packets and the policy to select the next packet in each queue add another layer of complexity to the determination of the \gls{aoi}. 
To the best of our knowledge, this is the first work to get a very good approximation and upper and lower bounds on the average \gls{aoi} and the total delay for general network topologies.
Moreover, we also derive an upper bound on the tail of the \gls{paoi}, which is critical for worst-case performance analysis.

The rest of the paper is organized as follows: Sec. \ref{sec:relatedwork} presents an overview of related work and the main contributions. Then, Sec. \ref{sec:systemmodel} describes the system model and the analytical tools to tackle the problem. Sec. \ref{sec:timedomain} presents the \gls{aoi} analysis for the general $K$-node case, with tight higher and lower bounds, while we derive an upper bound on the distribution tail in Sec. \ref{sec:tail}. Numerical results are presented in Sec. \ref{sec:results}. Finally, Sec. \ref{sec:conclusions} discusses the conclusions and future directions of this work. 

\section{Related work and contributions} \label{sec:relatedwork}

Latency has been widely studied in the context of satellite communications, from \gls{geo} to \gls{leo} orbits. However, the metric of interest has historically been the \gls{e2e} latency, defined as the time it takes a bit of information to traverse a network from its originating point to its final destination. Indeed, the \gls{e2e} delay performance was already investigated more than twenty years ago \cite{werner1997atm,Goyal1998} for a \gls{leo} satellite-ATM network. More recently,  \cite{Bisu2018} proposes a method for obtaining the \gls{e2e} latency in satellite IP-based networks and it is found that for \gls{geo} at least 50\% of the \gls{e2e} latency is due to the processing and transmission times. A tandem queue similar to ours is used in \cite{Wang2017stochastic} to model a multi-layered satellite network combining \gls{leo} and \gls{meo}. Specifically, stochastic network calculus is used to obtain the \gls{e2e} delay and backlog bounds. The problem of stochastic network calculus is, however, that the bounds can be very loose \cite{Wu2010}. The author in \cite{Handley2018} investigates the potential of one of the imminent commercial constellations, Starlink, to provide low-latency. In \cite{Dudukovich2019} a satellite relay network is considered, and \gls{ml}-based protocols for Delay Tolerant Networking are discussed. To the best of our knowledge, our previous paper \cite{Soret20202} was the first one addressing \gls{aoi} in a satellite set-up with inter-connected nodes, and in that prior work we provided initial results of the age in a multi-hop line system with ideal transmissions and \gls{fcfs} policy. 

As already mentioned, age-sensitive applications are those where a source generates updates that are transmitted through a communication network, like common satellite services that involve tracking processes or objects such as containers in logistics. 
The previously mentioned \gls{vdes} \cite{ITU2371}\cite{Lazaro2019} and AIS are meant to allow vessels to periodically report their position, course and speed, for collision avoidance, but the small assigned bandwidth makes the system design and the performance guarantee highly challenging. Another example is the \gls{adsb} system in airplanes \cite{ADSB}, where maintaining fresh information on the sender's status is the first and foremost objective of the network. 

The \gls{aoi} metric has been extensively studied in several different queueing systems: the original paper that defined it \cite{kaul2012real} analyzed the $G/G/1$ queue, concentrating on the exponential and deterministic distributions as case studies. Later works tried to calculate the \gls{aoi} and \gls{paoi} in specific realistic models of wireless channels, including errors~\cite{chen2016error} and retransmissions~\cite{devassy2019reliable} and verifying the queueing models with live experiments~\cite{beytur2019measuring}.  An interesting addition to the model is the consideration of multiple sources, which leads to a scheduling problem with the objective of limiting the age for each source~\cite{kadota2019minimizing} focused on the optimal scheduling protocols to improve the freshness of information in wireless networks. Optimizing the senders' updating policies in complex wireless communication systems has been proven to be an NP-hard problem, but near-optimal solutions can be achieved using greedy heuristics \cite{sun2017update} such as ``lazy updates'': each source can decide not to send some packet, waiting for new information to avoid overloading the queue with packets that give a limited benefit to the overall \gls{aoi} \cite{yates2015lazy}. Another possibility is to consider a limited transmission window for packets, after which they are dropped: in~\cite{li2020age}, the average \gls{paoi} is derived in such a scenario. Closed form expressions for the average \gls{aoi} of slotted and unslotted ALOHA have been given in~\cite{yates2017unreliable} and \cite{yates2020unslotted}. The metric has been compared to the performance of scheduled multiple access in~\cite{talak2018distributed}. Some selective packet transmission policies at source have been considered in~\cite{chen2020rach}, an optimal \gls{aoi} of a stabilized slotted ALOHA can be approximated by $1/N \lambda_0$ where $N \lambda_0$ is the sum arrival rate from N sources of updates if $N \rightarrow \infty$ and $N \lambda_0 < 1/e$.

If the connection is not single-hop, and there are several queueing systems in sequence, the network can be modeled as a tandem queue. Given the wide range of relevant applications, we focus our attention in the study of the age in this kind of models, which is particularly interesting in satellite relay applications, as \gls{leo}, but also the other, satellite networks are inherently multi-hop. If a single satellite relay is employed, the model is a 2-hop tandem queue, for which the average \gls{paoi} under \gls{fcfs} queueing was derived in~\cite{xu2019optimizing} for the multiple source case under $M/M/1$ systems. The Chernoff bound can be used to get an upper bound of the \gls{cdf} of the \gls{aoi} in these kinds of systems~\cite{champati2019statistical}, while our recent work~\cite{chiariotti2020peak} derives the distribution analytically.  

A general result was proven for queueing networks with any number of systems in~\cite{bedewy2017age} and ~\cite{bedewy2019multihop}: in any tandem of $M/M/1$ systems with a single source, the \gls{aoi} is minimized by applying the preemptive \gls{lcfs} policy. A more general result was derived in \cite{farazi2019bounds}, in which the authors study \gls{aoi} in a general multi-source multi-hop wireless network with explicit channel contention and have obtained upper and lower bounds for the average \gls{aoi} and \gls{paoi} based on fundamental graph parameters such as the connected domination number and average shortest path length. In \cite{talak2017minimizing}, the problem of multi-hop networks with many source-destination pairs and interference constraints is addressed, and the optimal policy is reduced to solving the equivalent problem in which all source-destination pairs are just a single-hop away. Queueing is a major source of delay, and updates do not need to be transmitted reliably, so it is often better to drop the packet in service and transmit the freshest one directly. A similar result has been proven for $M/M/k$ queues~\cite{bedewy2019minimizing}, and \cite{kam2018age} derives the average \gls{aoi} with preemption for 2-hop systems with different arrival processes. The problem is more interesting for different service time distributions, as the decision over whether to preempt or not becomes more complex~\cite{wang2018skip}. An analysis of the effect of preemption on tandem models on the average \gls{aoi} is presented in~\cite{yates2018age}: the work extends the stochastic hybrid system analysis, generalizing it for the moment generation function of the ageing process for a class of queueing networks with preemptive services and memoryless service times. Another possibility is queue replacement, in which only the freshest update for each source is kept in the queue, reducing queue size significantly: the replaced packet is not placed in the queue, but dropped altogether, reducing channel usage with respect to simple \gls{lcfs}, with or without preemption. In this case, the queue is modeled as an $M/M/1/2$, and if a new packet arrives it takes the queued packet's place. Some preliminary results on such a system are given in~\cite{pappas2015age}, while the average \gls{aoi} and \gls{paoi} are computed in~\cite{kosta2019queue} for one source and in~\cite{kosta2019age} for multiple sources. Finally, a general transport protocol to control the generation rate of status updates to minimize the \gls{aoi} over the Internet is presented in~\cite{shreedhar2019age}.

Our work considers a general queueing network with $K$ nodes and multiple sources and destinations, on which very little work has been done. Specifically, the main contributions of the paper are:
\begin{itemize}
    \item We model a satellite relay system as a multi-hop mesh network with packet losses and multiple sources and destinations. Each node $k$ receives traffic from multiple sources and forwards it to other satellites through the \gls{isl}, or to its destination through the downlink. This very general model is meant to represent satellite relay systems, but it can be applied to other multi-hop \emph{ad hoc} networks with multiple relays and multiple sink nodes, independently of their topology. An initial version of this model was presented in~\cite{Soret20202}, which analyzed a simple tandem queue with 2 satellite nodes. In this work, the model is fully general, and can represent any network with Poisson traffic and service, with arbitrary error rates for each link.
    \item We provide novel analytical results of the \gls{aoi} in this scenario: (1) a tight approximation and higher and lower bounds of the average \gls{aoi}; (2) an upper bound of the \gls{paoi} distribution; (3) and the exact value of the mean system delay. We show the results for a line network in which all sources have the same sink, and in a traditional dumbbell topology in which multiple connections share a single \gls{isl} link. The analysis is done for infinite buffers at each node, but it has been observed that having a limited storage capacity has little impact in the age performance of the multi-hop network. 
    \item We investigate user fairness and the impact of age-aware scheduling policies through the analysis of three queueing policies: \gls{fcfs}, \gls{opf} and \gls{haf}. The conventional \gls{fcfs} is a scheduling strategy well-suited for single queues, but it does not take into account the multi-hop nature of the network. Instead, our results show that the \gls{opf} and \gls{haf} policies, especially \gls{opf}, are able to increase the fairness between sources in different parts of the network while maintaining a similar average \gls{aoi}.
    \item We analyze the presence of wireless channel errors and their impact in the \gls{aoi}. A packet erasure channel models the losses in the wireless links. Lost packets are detrimental for the packet delay performance but, interestingly, the \gls{aoi} benefits from the load reduction of packet dropping when the system works close to congestion. This trade-off is yet another instance of the age-dilemma \emph{``how often should one update?''} \cite{kaul2012real}, although the answer in a system with multiple sources and multiple destinations is not trivial.  
\end{itemize}

\section{System model} \label{sec:systemmodel}

We consider a connection composed of $K$ links in a multi-hop mesh network in which each node in the network acts both as a source and a relay, as shown in Fig. \ref{fig_scenario}. The source is modeled as a Poisson process, generating packets with rate $\lambda$. Each node $k$ in the connection, including the source, receives Poisson cross traffic with rate $\theta_k$. Cross traffic might share part of the path with the packets from the source, and this is accounted for by the parameter $\psi_k$: a fraction $\psi_k$ of the cross traffic entering node $k$ leaves the connection, as it is transmitted through another link to nodes outside the considered source's path; the rest of the traffic is transmitted through the same path as the considered source's packets. In any case, the destination is on ground, for which the last node in the source's path represents the \gls{dl}. We assume that each satellite can receive and transmit packets at the same time over different \glspl{isl}. The \gls{isl} connects satellites in the same orbital plane or in different orbital planes, for which dedicated antennas are typically located in the roll and pitch axes, respectively. Moreover, the \gls{dl} has another dedicated antenna pointing at the Earth's center. If the information from the source is delay-sensitive (e.g., status updates), it must be routed as soon as possible to the destination, and \gls{aoi} is a useful metric for system performance.

\begin{figure}[t]
	\centering
	\includegraphics[width=6.5in]{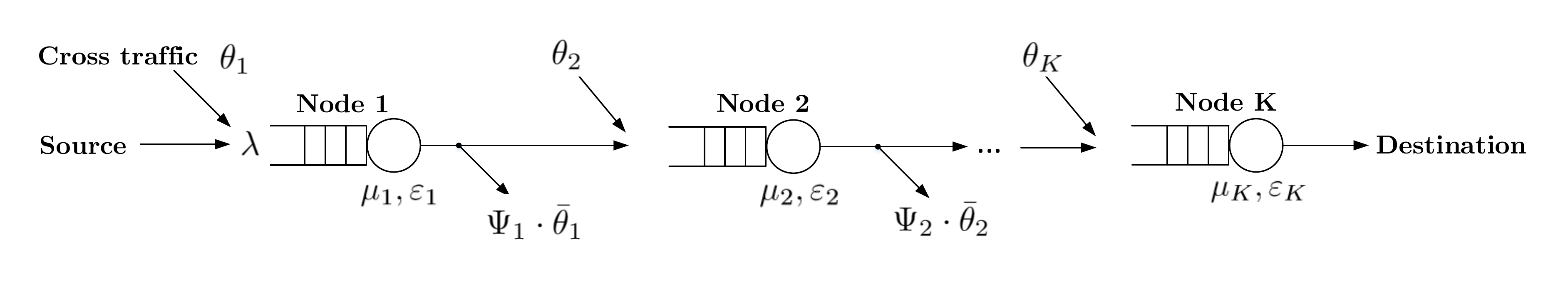}
	\caption{General multi-hop queueing system with cross traffic.}
	\label{fig_system_model}
\end{figure}

We model a connection between a source and a destination as an $M/M/1$ queueing network connected in series. In a real system, the service time for each link depends on the length of the packet and the quality of the link: in this work, we model the service time for each link for the same packet as independent for tractability. This assumption is equivalent to considering uncorrelated distances between pairs of satellites; considering a correlated system is left for future work.
Each node $k$ receives traffic from the node, as well as cross traffic, some of which is then routed through other connections or arrives at its destination. This model is fully general, as it can describe any network with Poisson sources, from the point of view of any of the traffic flows in the network. We model the $k$-th link in the connection, between nodes $k$ and $k+1$, as an erasure channel with an error probability $\varepsilon_k$: any packet sent by node $k$ is correctly received by node $k+1$ with probability $1-\varepsilon_k$. The service follows a Poisson process with rate $\mu_k$, i.e., the average service time of each link is the inverse of the service rate, $S_k = 1/\mu_k$. This model is general enough to capture heterogeneous capabilities and losses in a satellite network, where the links (\gls{isl} in the same orbital plane or between different planes, and the \gls{dl}) can be of a highly different nature. The system can be entirely described by the source rate $\lambda$ and the vectors $\bm{\theta}$, $\bm{\psi}$, $\bm{\mu}$, and $\bm{\varepsilon}$, which describe the cross traffic and the channels' statistical properties.
Using these vectors, we can compute the total cross traffic load at node $k$, denoted as $\bar{\theta}_k$:
\begin{equation}
    \bar{\theta}_k=\sum_{j=1}^k \theta_j \prod_{i=j}^{k-1}(1-\psi_i)(1-\varepsilon_i).
\end{equation}
As Fig.~\ref{fig_system_model} shows, each node in the connection receives traffic from the previous node, as well as cross traffic: a part of the cross traffic leaves the connection, as it is transmitted through other paths, while the rest is transmitted along the connection with the considered source's packets.

\begin{figure}[!t]
 \centering
 \includegraphics{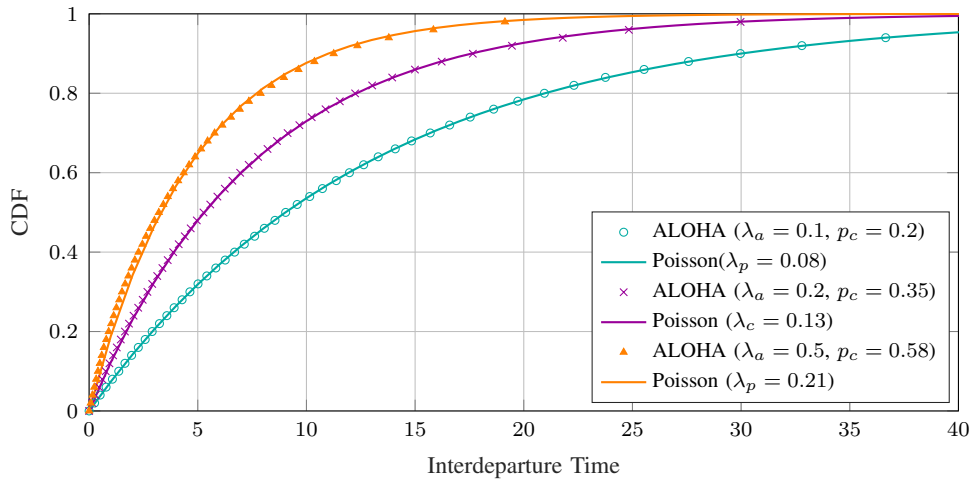}
 \caption{CDF of the interdeparture time.}
 \label{fig:departure_cdf}
\end{figure}

The arrival process at the source node models the \gls{ul} access in the satellite network. This access can be implemented in many different ways. For a large number of intermittent transmitters in a single shared communication channel, the ALOHA protocol is the simplest one and it is widely used \cite{Norman1970}. Rather than the conventional ALOHA implementation with a backoff mechanism to re-send colliding packets, a pure ALOHA scheme with a single transmission attempt is more suitable for age-sensitive applications. Thus, the source transmits each new available status update immediately and a collision occurs when other users transmit their packets simultaneously. We can consider two extreme cases in this part of the model:
\begin{itemize}
    \item \emph{Ideal \gls{mpr}}. In this case~\cite{Goseling2015} the packets are not lost due to collisions and can only be lost due to channel errors. This model is suitable for IoT systems based on \gls{unb} transmissions, such as SigFox \cite{mroue2018mac}, where the receiver is designed to take advantage of the very small bandwidth occupied by a single packet and decode multiple packets simultaneously.
    \item \emph{Destructive collisions.} The other extreme is adoption of the classical ALOHA model, in which any collision is destructive and all packets involved in the collision are lost. Strictly speaking, here the resulting process is not Poisson due to the correlation created when multiple packets are lost in a collision create.
\end{itemize}
Regardless of whether we model losses as destructive collisions or channel errors, the failed sources will not try again, but just wait until the next status update is generated, and $p_c$ is the probability of incorrect packet decoding due to either channel error or collision.
In case of \gls{mpr}, the arrival process of the packets that reach the first satellite is still Poisson, but thinned with  probability $(1-p_c)$, such that the resulting arrival rate is $\lambda (1-p_c)$. In case of 
destructive collisions, the error source is both channel noise and collision. For this case the thinned Poisson process with arrival rate $\lambda (1-p_c)$ is only an approximation. As shown in Fig. \ref{fig:departure_cdf}, where we compare the ALOHA departure process with arrival rate $\lambda_a$ and a Poisson arrival process with rate $\lambda_p=\lambda_a(1-p_c)$, the approximation is justified for a wide range of arrival rates. This will be verified in the overall results as well.

With this model, the difference between the age before the access and the age at the queue of the first satellite is just a constant delay corresponding to the \gls{ul} transmission time.  We denote the \gls{aoi} in the destination node at time $t$ as the random process $\Delta(t)$, which increases linearly in time in the absence of any updates, and is reset to a smaller value when an update is received. For the reader's convenience the complete list of notations is given in Table~\ref{tab:Notations}. In addition, for a random variable (RV) $X$, $\mathbb{E}[X]$ stands for the expected value and $f_{X}(x)$ denotes Probability Distribution Function (PDF) of $X$. 
 
Note that, if we use the standard \gls{fcfs} policy, traffic from all sources is stored in the same buffer, with no priorities among them. We will later analyze the case in which nodes apply the \gls{opf} and \gls{haf} queueing policies. \gls{opf} prioritizes packets by their generation time instead of their arrival time at the node, enhancing fairness between different flows, as packets that have already gone through longer connections can traverse later links faster, at the expense of fresher packets from sources closer to their destination. Differently, \gls{haf} is not aimed at fairness, but at improving \gls{aoi}, as it prioritizes packets whose source has the highest current \gls{aoi} at the node.
 
 \begin{table}[t]
     \centering
     \footnotesize
     \caption{Relevant notation}\vspace{0.25cm}
     \label{tab:Notations}
     \begin{tabular}{p{1cm}p{6cm}|p{1cm}p{6cm}} 
          \hline
          Notation & Definition & Notation & Definition \\ 
          \hline \hline
          $K$ & Number of links in a multi-hop network & $N(\mathcal{T})$ & Number of arrivals from source by time $\mathcal{T}$ \\
          $\lambda$ & Packets generation rate at source & $\Delta_{\mathcal{T}}$ & Time average \gls{aoi} over $\mathcal{T}$ \\
          $\theta_k$ & cross traffic rate at node $k$ &$\bar{\Delta}$ & Average \gls{aoi} \\
          $\bar{\theta}_k$ & Total cross traffic load at node $k$ & $\rho_k$ & Traffic load at node $k$ \\
          $\bm{\theta}$ & Vector of cross traffic rates & $\rho$ & Error free load \\
          $\psi_k$ & Probability of cross traffic offloading in $k$ & $S_k$ & Average service time at node $k$ \\
          $\bm{\psi}$ & Vector of cross traffic offloading probabilities & $\Delta(t)$ & Ageing process \\
          $p_s(j)$ & Packet delivery success probability over $j$ links & $\xi_i$ & \gls{paoi} of packet $i$ \\
          $\varepsilon_k$ & Channel error probability for the $k$-th link & $\alpha_k$ & Packets response rate at node $k$ \\
          $\bm{\varepsilon}$ & Vector of channel error probabilities & $\bm{\alpha}$ & Vector of response rates \\
          $\mu_k$ & Packet service rate at node $k$ & $t_i$ & Status update $i$ generation time (by source) \\
          $\bm{\mu}$ & Vector of service rates & $t'_i$ & Status update $i$ time at monitor (on ground) \\ 
          $\delta_{ij}$ & Hypoexponential distribution coefficient of packet service time & $\Pi_j(n)$ & Steady-state distribution of the number of queued packets at node $j$ \\
          $\gamma_{ij}$ & Hypoexponential distribution coefficient of packet total network time & $\bm{\omega}$ & Vector of hypoexponentional distribution parameters for the \gls{paoi} bound \\
          $Y_i$ & Packet interarrival time  & $p_c$ & Uplink collision probability\\ 
          $Z_i$ & Packet interdeparture time & $Q_{i}$ & Area under the $\Delta(t)$ \gls{aoi} process\\
          $T_i$ & Packet $i$ network time &  $Q'_{i}$ & Additional area below the $\Delta(t)$ process after a  \\
          $T_{i,j}$ & Packet $i$ system time at node $j$  & & missed packet\\
          $W_i$ & Packet $i$ total waiting time & $Q^{(n)}_{i}$ & Total area around $\Delta(t)$ process after $n$ missed  \\
          $W_{i,j}$ & Packet $i$ waiting time at node $j$ & & packets\\
          $S_i$ & Packet $i$ total service time & $\Omega_j$ & Time difference between arrival and departure   \\
         $S_{i,j}$ & Packet service time at node $j$ & & time of two consecutive packets at node $j$ \\
          \hline
     \end{tabular}
 \end{table}
 
 \subsection{Average AoI in the error-free scenario}
 
 \begin{figure}[!t]
 \centering
 \includegraphics{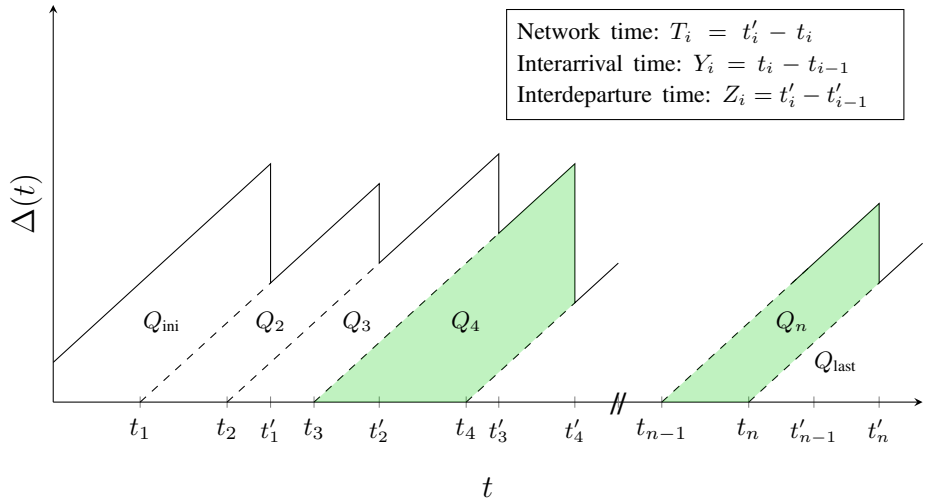}
 \caption{Evolution of the Age of Information in a queue network with $K$ nodes. The network times $T_i$ are defined as the total time spent in the system, since arrival in node $1$ until departure in node $K$.}
 \label{fig:aoi_FCFS}
\end{figure}
 
 We first consider an error-free scenario, in which $\varepsilon_k=0\;\forall k$. In this case, the evolution of the \gls{aoi} $\Delta(t)$ for source under a \gls{fcfs} policy exhibits the sawtooth pattern plotted in Fig. \ref{fig:aoi_FCFS}. Without loss of generality, the system is first observed at $t=0$ and the queue is empty with age $\Delta(0)$. The status update $i$ is generated at time $t_i$ and is received by the ground station at time $t_i'$. We define $Y_i$ as the interarrival time $Y_i = t_i - t_{i-1}$ between two packets, $Z_i$ as the interdeparture time $Z_i = t'_i - t'_{i-1}$, and $T_i$ as the total network time in the system $T_i = t'_i - t_i$. The latter includes the time spent in all the nodes (queueing time and transmission time) until departure from the system at node $K$. Our definitions follow the work in \cite{kuang2019age}, which considered a single buffered system, but in our case, the \gls{e2e} connection is modeled as a sequence of $M/M/1$ systems, each of which has to deal with cross traffic. 

To evaluate the average \gls{aoi}, the strategy is to calculate the area under $\Delta(t)$, or the time average \gls{aoi}, as
\begin{equation}
    \Delta_{\mathcal{T}} = \frac{1}{\mathcal{T}} \left(Q_{\text{ini}} + Q_{\text{last}}+\sum_{i=2}^{N(\mathcal{T})} Q_i \right),
\end{equation}
where $N(\mathcal{T})$ is the number of arrivals from the source by time $\mathcal{T}$. The average \gls{aoi} $\bar{\Delta}$ is given by the limit
$\bar{\Delta} = \lim_{\mathcal{T} \rightarrow \infty} \Delta_{\mathcal{T}} $.
As defined in Fig. \ref{fig:aoi_FCFS}, each $Q_i$ (with $i>1$) is a trapezoid whose area can be calculated as the difference between two isosceles triangles \cite{kuang2019age}, i.e., 
\begin{align}
Q_i = \frac{1}{2}\left(T_i + Y_i\right)^2 - \frac{1}{2}T_i^2 = Y_i T_i + \frac{Y_i^2}{2}.
\end{align}
The average \gls{aoi} $\bar{\Delta}$ can then be expressed as
\begin{align}
\bar{\Delta} &= \lambda \mathbb{E}\left[Q_i\right] =
\lambda \left( \mathbb{E}\left[T_i Y_{i}\right] +
\mathbb{E}\left[\frac{1}{2}Y_{i}^2\right]\right)' \label{eq:aoi}.
\end{align}

Ergodicity has been assumed for the stochastic process $\Delta(t)$, but no assumptions regarding the distribution of the random variables $Y$ and $T$ have been made. 
Considering that the arrival process is Poisson, the interarrival times $Y_i$ are exponentially distributed with rate $\lambda$. The second term in \eqref{eq:aoi}, $\mathbb{E}\left[\frac{1}{2}Y_{i}^2\right]$, is then easily obtained as $
\mathbb{E}\left[\frac{1}{2}Y_{i}^2\right] = \frac{1}{\lambda^2}$.
The other term in \eqref{eq:aoi} is harder to derive, as $T_i$ is correlated to $Y_i$. Intuitively, a packet coming right after another packet from the same source will experience a higher queueing delay, while a packet that arrives a long time after the previous one from the same source will just have to deal with the cross traffic, as all packets from the source will have already been transmitted.

\subsection{Average AoI in the error-prone scenario}

\begin{figure}[!t]
 \centering
 \includegraphics{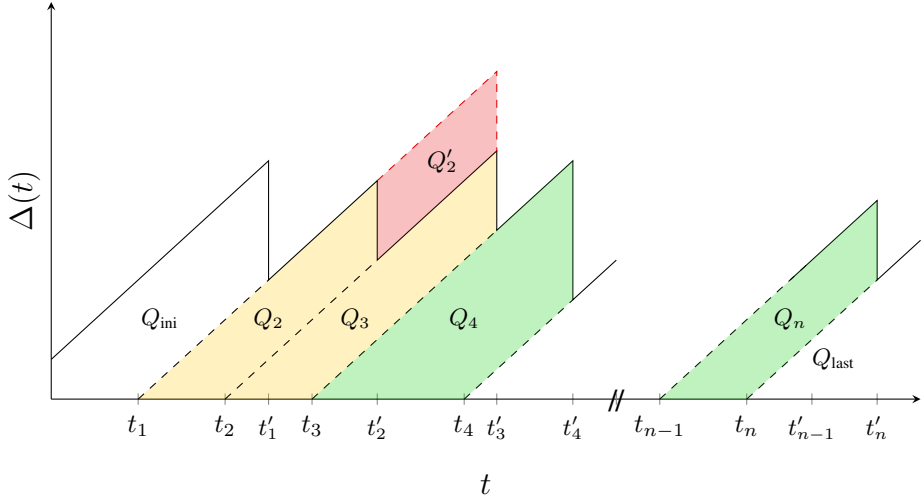}
 \caption{Evolution of the Age of information in a queue network with $K$ nodes and errors. The additional area $Q'_2$, highlighted in red, shows the increase in the \gls{aoi} in case a packet is lost.}
 \label{fig:aoi_error}
\end{figure}

We now consider the more general case with transmission errors: we assume that each link $j$ has a known error probability $\varepsilon_j$, and that a lost packet has the same service time as a correctly received packet, but is not put in the queue for the next node. Fig.~\ref{fig:aoi_error} shows the geometric analysis for the evolution of the \gls{aoi} for a specific source in case of errors: if packet 2 is dropped, the additional area $Q'_2$, highlighted in red in the figure, needs to be included in the computation. If we consider the $Q^{(1)}_i=Q_i+Q_{i-1}+Q'_{i-1}$ trapezoid resulting from a single lost packet, its area is given by:
\begin{align}
Q^{(1)}_i=\frac{1}{2}\left(T_i+Y_i+Y_{i-1}\right)^2-\frac{1}{2}T_i^2=Y_i T_i +Y_{i-1} T_i +\frac{Y_i^2}{2}+\frac{Y_{i-1}^2}{2}.
\end{align}
The trapezoid is the sum of the red and yellow areas in the figure. We can generalize this result to the case with $n$ errors:
\begin{align}
Q^{(n)}_i&=\sum_{j=0}^n \left[ Y_{i-j}T_i+\frac{1}{2}Y_{i-j}^2 \right]+\sum_{j=0}^n\sum_{\ell=0,\ell\neq j}^n Y_{i-j}Y_{i-\ell}
\end{align}
We note here that, since each node is an $M/M/1$ queue, the interarrival times $Y_i$ and $Y_{i-j}$ are independent for any $j>0$, and so are $T_i$ and $Y_{i-j}$. When the system reaches a steady state, the system times are stochastically identical, i.e., $T = ^{st} T_i =^{st} T_{i-1}$, and the same holds for the interarrival times. Since errors are assumed to be independent, and the probability of delivering a packet through the first $j$ links correctly is $p_s(j)=\prod_{i=1}^j \left(1-\varepsilon_i\right)$, the average \gls{aoi} is given by:
\begin{equation}
    \begin{aligned}
    \bar{\Delta}&=\lambda\sum_{n=0}^\infty p_s(K) \left(1-p_s(K)\right)^n\mathbb{E}\left[Q^{(n)}_i\right]\\
            &=\lambda\sum_{n=0}^\infty p_s(K) \left(1-p_s(K)\right)^n\left(\mathbb{E}\left[Y_i T_i\right]+n\mathbb{E}\left[Y_{i-1}T_i\right]+\frac{n+1}{2}\mathbb{E}\left[Y_i^2\right]+\binom{n}{2}\mathbb{E}\left[Y_i\right]^2\right).
    \end{aligned}\label{eq:aoi_err_partial}
\end{equation}
The total arrival rate at each node $j$ is given by the surviving packets from the source and the cross traffic, and it is $p_s(j)\lambda+\bar{\theta}_j$. We then define the response rate at node $j$ as $\alpha_j=\mu_j-(p_s(j)\lambda+\bar{\theta}_j)$. If all satellites apply the \gls{fcfs} queueing policy, the total system time in the $j$-th node in steady-state is a Poisson process with rate $\alpha_j$, according to Little's law. The overall service time and waiting time of the connection then follow a Hypoexponential distribution~\cite{amari1997closed}. 

The vector $\bm{\alpha}$, containing the response rates for the $K$ links, has $N$ unique elements, each appearing $n_i$ times. If $N=1$ and $n_1=K$, all rates are the same and the total system time follows an Erlang distribution. In fact, the Hypoexponential distribution is the convolution of several Erlang distributions.
The density function $f_T(t)$ of the total network time is given in~\cite{jasiulewicz2003convolutions} as:
\begin{align}
f_T(t)&=\sum_{i=1}^{N}\sum_{j=1}^{n_i} \gamma_{ij}\frac{t^{j-1}}{(j-1)!}e^{-\alpha_i t},\label{eq:hypo}
\end{align}
where $\gamma_{ij}$ is a coefficient defined as:
\begin{align}
\gamma_{ij}=\prod_{\ell=1}^{N}\left(\alpha_\ell^{n_\ell}\right)(-1)^{n_i-j}\sum_{\sum_{\ell=1}^{N} m_\ell=n_i-j, m_i=0}\prod_{\ell=1,\ell\neq i}^{N} \binom{n_\ell+m_\ell-1}{m_\ell}\frac{1}{(\alpha_\ell-\alpha_i)^{n_\ell+m_\ell}}.\label{eq:hypo_coeff}
\end{align}
Since the sum in~\eqref{eq:aoi_err_partial} converges for $p_s(K)\in(0,1]$, we can then exploit the properties of the Hypoexponential distribution to get:
\begin{equation}
    \begin{aligned}
      \bar{\Delta}&=\lambda \left(\mathbb{E}\left[Y_i T_i\right]+\frac{1-p_s(K)}{p_s(K)}\mathbb{E}\left[Y_{i-1}\right]\mathbb{E}\left[T_i\right]+\frac{1}{2 p_s(K)}\mathbb{E}\left[Y_i^2\right]+\left(\frac{1-p_s(K)}{p_s(K)}\right)^2\mathbb{E}\left[Y_i\right]^2\right)\\
      &=\lambda \left(\mathbb{E}\left[Y_i T_i\right]+\sum_{j=1}^{N}\frac{1-p_s(K)}{p_s(K)\alpha_j^{n_j}\lambda}+\frac{1}{\lambda^2 p_s(K)}+\left(\frac{1-p_s(K)}{\lambda p_s(K)}\right)^2\right).\label{eq:aoi_err}
    \end{aligned}
\end{equation}
The $\mathbb{E}\left[Y_i T_i\right]$ term is complex, as it depends on the correlation between interarrival time and subsequent system time: its analytical derivation is too cumbersome to calculate for the general case, but we can find lower and upper bounds on the average \gls{aoi} for each source.

\section{Age of Information bounds and approximation} \label{sec:timedomain}

In this section, we start from the result in~\eqref{eq:aoi_err} and derive lower and upper bounds for the average \gls{aoi} for the \gls{fcfs}, \gls{opf}, and \gls{haf} policies as well as a reasonably tight approximation. In the system we consider, the distribution of the total system time is the one given in~\eqref{eq:hypo}; using the independence assumption for service times, the distribution of the service time $f_S(t)$ is another Hypoexponential, using $\mu_i$ instead of $\alpha_i$:
\begin{equation}
    f_S(t)=\sum_{i=1}^{N'}\sum_{j=1}^{n'_i} \delta_{ij}\frac{t^{j-1}}{(j-1)!}e^{-\mu_i t},\label{eq:hypo2}
\end{equation}
We denote the Hypoexponential coefficients for the service time distribution, computed by applying \eqref{eq:hypo_coeff} using the service rate vector instead of $\bm{\alpha}$, as $\delta_{ij}$, to avoid confusion. $N'$ and $n_i'$ are the equivalents of $N$ and $n_i$ for the vector $\bm{\mu}$.

In order to find the average \gls{aoi}, we need to compute the first term in~\eqref{eq:aoi_err}, $\mathbb{E}\left[T_i Y_{i}\right]$. The total system time $T_i$ of packet $i$ is the sum of the system times in each of the nodes $1,2,\ldots,K$, and each of them can be decomposed in waiting and service time:
\begin{equation}
T_i = W_{i,1} + S_{i,1} + W_{i,2} + S_{i,2} +\ldots + W_{i,K} + S_{i,K}.
\end{equation}
We rewrite this term to get:
\begin{align}
\mathbb{E}\left[T_iY_{i}\right] &= \mathbb{E}\left[(W_i + S_i) Y_{i}\right] = \mathbb{E}\left[W_i Y_{i}\right] + \mathbb{E}\left[ S_i \right]\mathbb{E}\left[Y_{i}\right] \nonumber \\ &= \mathbb{E}\left[W_i Y_{i}\right] + \sum_{j=k}^{K}\mathbb{E}\left[ S_{i,j}\right]\mathbb{E}\left[Y_{i}\right] \label{eq:tiyi}.
\end{align}
Service times are independent from interarrival times, so we can simplify the second term, but $\mathbb{E}\left[W_i Y_{i}\right]$ is still complex.

First, we can consider a simple approximation by making the strong assumption that the interarrival and waiting times are independent, getting $\mathbb{E}\left[W_i Y_i\right] \simeq \mathbb{E}\left[W_i\right]\mathbb{E}\left[Y_i\right]$. The average \gls{aoi} is then approximated by:
\begin{equation}
    \begin{aligned}
      \bar{\Delta}&\simeq\sum_{j=1}^{N}\frac{1}{p_s(K)\alpha_j^{n_j}}+\frac{1}{\lambda p_s(K)}+\frac{(1-p_s(K))^2}{\lambda p_s(K)^2}.\label{eq:aoi_approx}
    \end{aligned}
\end{equation}
In the following, we will derive explicit upper and lower bounds by finding easily computable bounding random variables for each of the policies we consider.

\subsection{Bounds on the average \gls{aoi} for the \gls{fcfs} policy}

The total waiting time in the network $W_i$ is given by $W_{i,1}+\ldots+W_{i,K}$. The waiting time at each node depends on the time difference between the arrival of the new packet and the departure of the previous packet from the node. This time difference, which we denote as $\Omega_j$, is given by:
\begin{align}
  \Omega_j=\begin{cases}
             T_{i-1,1}-Y_i &\text{if } j=1;\\
             T_{i-1,j}-W_{i,j-1}-S_{i,j-1}&\text{if } j>1.
           \end{cases}
\end{align}
The total waiting time is then simply given by:
\begin{equation}
    \begin{aligned}
  W_i&=\sum_{j=1}^K(\Omega_j)^+=\left(T_{i-1,1}-Y_i\right)^+ +\sum_{j=2}^K\left(T_{i-1,j}-W_{i,j-1}-S_{i,j-1}\right)^+,
    \end{aligned}
\end{equation}
where $(x)^+=\max(x,0)$ is the positive part function. It is trivial to prove that the sum of positive parts is larger than the positive part of the sum:
\begin{equation}
  \sum_{i=1}^n (x_i)^+ \geq \left(\sum_{i=1}^n x_i\right)^+ \forall \mathbf{x}\in\mathbb{R}^n, \forall n\in\mathbb{N}.
\end{equation}
Thus, we can write a lower bound on the total waiting time of packet $i$ in the general case of $K$ nodes as:  
\begin{equation}
    \begin{aligned}
W_{i}\geq \left(\sum_{j=1}^K\Omega_j \right)^+= \left(T_{i-1} - Y_{i} - \sum_{j=1}^{K-1}S_{i,j}\right)^+ = \left(T_{i-1} - Y_{i} - S_{\setminus K}\right)^+,
\label{eq:wi}
    \end{aligned}
\end{equation}
where we have defined $S_{\setminus K}=\sum_{j=1}^{K-1}S_{i,j}$. Note also that the bound in \eqref{eq:wi} becomes equality if the packet is queued at each node, i.e., it never finds an empty queue. In the case in which one or more of the queues are empty, the time between the departure of packet $i-1$ and the arrival of packet $i$ should be added to the result above. 
We can now write the lower bound on the conditional expected waiting time:
\begin{equation}
    \begin{aligned}
\mathbb{E}\left[W_i | Y_i = y, S_{\setminus K}=s\right] &\geq \mathbb{E}\left[(T-y-s)^+\right]. \label{eq:conditional_nointer}
    \end{aligned}
\end{equation}
To solve \eqref{eq:conditional_nointer}, we use the distribution of the system time. If the system is not saturated, i.e., if the server utilization in each $M/M/1$ stage meets the stability condition $\rho_j = \frac{p_s(j)\lambda+\bar{\theta}_j}{\mu_j} < 1$, then Burke's theorem can be applied \cite{burke1956output}. This means that the departure process from each node $j$ is also a Poisson process, and each node can be analyzed separately. First, we derive the lower bound on the conditioned expectation $\mathbb{E}\left[W_i Y_i | S_{\setminus K}=s\right]$:
\begin{equation}
    \begin{aligned}
    \mathbb{E}\left[W_i Y_i | S_{\setminus K}=s\right]\geq&\int_0^{\infty}y \mathbb{E}\left[W_i|Y_i=y,S_{\setminus K}=s\right] f_{Y_i}(y) dy\\
    \geq&\int_0^{\infty}y\int_{y+s}^{\infty} (t-y-s) f_T(t) f_{Y_i}(y) dt dy\\
    \geq&\int_0^{\infty}\sum_{i=1}^{N}\sum_{j=1}^{n_i}\sum_{\ell=0}^j\frac{y\lambda e^{-\lambda y}\gamma_{ij}e^{-\alpha_i(y+s)} (y+s)^\ell \left(j-\ell\right)}{(\ell!)\alpha_i^{j-\ell+1}} dy\\
    \geq&\sum_{i=1}^{N}\sum_{j=1}^{n_i}\sum_{\ell=0}^j \frac{\gamma_{ij}\lambda\left(j-\ell\right)}{(\ell!)\alpha_i^{j-\ell+1}} e^{-\alpha_i s}\int_0^{\infty} y(y+s)^\ell e^{-(\alpha_i+\lambda) y} dy\\
    \geq&\sum_{i=1}^{N}\sum_{j=1}^{n_i}\sum_{\ell=0}^{j}\sum_{m=0}^{\ell+1}\frac{\lambda\gamma_{ij}(j-\ell)(\ell-m+1) s^m e^{-\alpha_i s}}{(m!)\alpha_i^{j-\ell+1}\left(\alpha_i+\lambda\right)^{\ell-m+2}}.
    \end{aligned}
\end{equation}
We can now get the lower bound on the unconditioned expectation by applying the law of total probability again, knowing that the distribution of the service time in the first $K-1$ nodes is a Hypoexponential with parameter vector $\bm{\mu}_{\setminus K}=(\mu_1,\ldots,\mu_{K-1})$:
\begin{equation}
    \begin{aligned}
    \mathbb{E}\left[W_i Y_i\right]\geq&\int_0^{\infty} \mathbb{E}\left[W_i Y_i|S_{\setminus K}=s\right] f_{S_{\setminus K}}(s) ds\\
    \geq&\sum_{i=1}^{N}\sum_{j=1}^{n_i}\sum_{\ell=0}^j\sum_{m=0}^{\ell+1}\sum_{o=1}^{N'}\sum_{p=1}^{n'_\ell}\frac{\gamma_{ij}\delta_{\ell m}\lambda\left(j-\ell\right)\left(\ell-m+1\right)}
    {\alpha_i^{j-\ell+1}\left(\alpha_i+\lambda\right)^{\ell-m+2}m!(p-1)!}\int\displaylimits_0^{\infty} s^{m+p-1}e^{-(\alpha_i+\mu_o)s} ds\\
    \geq&\sum_{i=1}^{N}\sum_{j=1}^{n_i}\sum_{\ell=0}^j\sum_{m=0}^{\ell+1}\sum_{o=1}^{N'}\sum_{p=1}^{n'_\ell}\binom{m+p-1}{m}\frac{\gamma_{ij}\delta_{\ell m}\lambda\left(j-\ell\right)\left(\ell-m+1\right)}
    {\alpha_i^{j-\ell+1}\left(\alpha_i+\lambda\right)^{\ell-m+2}\left(\alpha_i+\mu_o\right)^{m+p}}\label{eq:lower},
    \end{aligned}
\end{equation}
where, as above, $N'$ and $n_i'$ are the equivalents of $N$ and $n_i$ for the vector $\bm{\mu}_{\setminus K}$: $N'$ is the number of unique elements of the vector and $n_i'$ is the multiplicity of the $i$-th such unique element.

We also give a simple formulation of an upper bound on the expected waiting time. We know that $W_{i,k}\leq T_{i-1,k}$, and that $T_{i-1,k}$ is independent from $Y_{i-1}$ for any value of $k$, so $\mathbb{E}\left[T_{i-1}Y_i\right]=\mathbb{E}\left[T_{i-1}\right]\mathbb{E}\left[Y_i\right]$. We can then exploit the fact that $\mathbb{E}\left[T_{i-1}Y_i\right]\geq \mathbb{E}\left[W_iY_i\right]$:
\begin{equation}
\begin{aligned}
    \mathbb{E}\left[W_iY_i\right]\leq&\mathbb{E}\left[T_{i-1}\right]\mathbb{E}\left[Y_i\right]\leq\sum_{j=1}^{N} \frac{1}{\lambda\alpha_j^{n_j}}\label{eq:upper},
\end{aligned}
\end{equation}
knowing that the average of a Hypoexponential distribution is the sum of the inverse of the rate of each link.

Now that we have derived bounds on $\mathbb{E}\left[W_i Y_i\right]$, we can derive the bounds for the average \gls{aoi}. The last term in (\ref{eq:tiyi}) is easily obtained as $\mathbb{E}\left[ S_i\right]\mathbb{E}\left[Y_{i}\right]=\sum_{j=1}^{N} \frac{1}{\lambda \mu_j^{n_j}}$. We can then obtain the bounds on $\bar{\Delta}$ by substituting the values of~\eqref{eq:lower} and~\eqref{eq:upper} into:
\begin{equation}
    \bar{\Delta}=\lambda\left(\mathbb{E}\left[W_i Y_i\right]+\sum_{j=1}^{N}\left( \frac{1}{\mu_j^{n_j}\lambda}+\frac{1-p_s(K)}{p_s(K)\alpha_j^{n_j}\lambda}\right)+\frac{1}{\lambda^2 p_s(K)}-\left(\frac{1-p_s(K)}{\lambda p_s(K)}\right)^2\right).\label{eq:bounds}
\end{equation}

\subsection{Bounds on the average \gls{aoi} for the \gls{opf} and \gls{haf} policies}\label{ssec:n_nodes_opf}

The \gls{opf} queueing policy is a simple twist on \gls{fcfs} that can improve the fairness among nodes: instead of using \gls{fcfs} at each node, packets are timestamped at the source, and each node transmits the packet in the queue with the lowest timestamp. In this way, packets generated farther away from their destinations do not have to wait in line at each node, but are given a higher priority if they have already spent more time in the network.
Using \gls{opf}, the \gls{fcfs} principle is not applied to each single node, but to the system as a whole: packets that are \emph{generated} first are served first, regardless of the order of arrival at each specific node in the tandem queue. The \gls{haf} policy extends the benefits of \gls{opf} by considering age explicitly: the source with the highest current age at the node is prioritized.

The lower bound is based on two simplifying assumptions, both of which reduce the age by removing possible cases from the calculation: firstly, we consider the waiting time when no queue is empty, i.e., when the transmission of a packet begins immediately after the previous packet is sent. Secondly, we only consider the case in which the packets that arrive at a certain node after the one we consider are all younger: in reality, cross traffic packets might be older than the one coming through the \gls{isl}, but this case complicates the analysis considerably, even though it gives a tighter bound. The bound is the same for both \gls{opf} and \gls{haf}, although the two sources choose the priorities of packets in slightly different ways.

We denote the steady-state distribution of the number of waiting packets at node $j$ as $\Pi_j(n)=(1-\rho_j)\rho_j^n$, reminding the reader that $\rho_j=\frac{p_s(j)\lambda+\bar{\theta}_j}{\mu_j}$ is the traffic load at node $j$. Let $L_j$ be the number of packets in a queue $j$. The conditioned expected waiting time for the $j$-th queue is always larger than:
\begin{equation}
\begin{aligned}
    \mathbb{E}\left[W_{i,j}|Y_i=y,S_{i,j-1}=s\right]\geq&\sum_{n=0}^\infty \Pi_j(n) \mathbb{E}\left[W_{i,j}|Y_i=y,S_{i,j-1}=s,L_j=n\right]\\
    \geq&\sum_{n=0}^\infty \left(1-\rho_j\right) \rho_j^n \int_s^\infty \frac{\mu_j^n t(t-s)^{n-1} e^{-\mu_j(t-s)}}{(n-1)!} dt\\
    \geq& (1-\rho_j)\rho_j s e^{-\alpha_j s}.
\end{aligned}
\end{equation}
We now apply the law of total probability, considering the nodes past the first one, i.e., for $j>1$:
\begin{equation}
\begin{aligned}
    \mathbb{E}\left[W_{i,j}|Y_i=y\right]\geq&\int_0^\infty P_{S_{j-1}}(s) \mathbb{E}\left[W_{i,j}|Y_i=y,S_{i,j-1}=s\right] ds\\
    \geq&\int_0^\infty \mu_{j-1} e^{-\mu_{j-1} s}(1-\rho_j)\rho_j s e^{-\alpha_j s} ds\\
    \geq& \frac{(1-\rho_j)\rho_j\mu_{j-1}}{\alpha_j+\mu_{j-1}}
\end{aligned}
\end{equation}
We now uncondition over $Y_i$, using the law of total probability:
\begin{equation}
  \mathbb{E}\left[W_{i,j}Y_i\right]\geq\frac{(1-\rho_j)\rho_j\mu_{j-1}}{(p_s(j)\lambda+\bar{\theta}_j)\left(\alpha_j+\mu_{j-1}\right)}.
\end{equation}
Naturally, since the first node is an $M/M/1$ queue with \gls{fcfs} queueing policy, the value of the expected queueing time $\mathbb{E}[W_{i,1}]$ is:
\begin{align}
    \mathbb{E}\left[W_{i,1}|Y_i=y\right]=&\frac{e^{-\alpha_1 y}}{\alpha_1}.
\end{align}
Unconditioning over $Y_i$, we get:
\begin{align}
    \mathbb{E}\left[W_{i,1}Y_i\right]=&\frac{\lambda}{\alpha_1\mu_1^2}.
\end{align}
The lower bound on the expected value $\mathbb{E}\left[W_iY_i\right]$ is then given by:
\begin{align}
    \mathbb{E}\left[W_i Y_i\right] \geq& \frac{\lambda}{\alpha_1\mu_1^2} + \sum_{j=2}^K \frac{(1-\rho_j)\rho_j\mu_{j-1}}{(p_s(j)\lambda+\bar{\theta}_j)\left(\alpha_j+\mu_{j-1}\right)}.\label{eq:opf_lower}
\end{align}
The upper bound on the \gls{aoi} is the same as for the \gls{fcfs} system. We then get the lower and upper bound by substituting~\eqref{eq:opf_lower} and~\eqref{eq:upper}, respectively,  into~\eqref{eq:bounds}.

\subsection{PAoI bound on the tail distribution in the error-free case}\label{sec:tail}

The average \gls{aoi} is an important parameter to design a tracking system, but information on the tail of the \gls{paoi} distribution is often required to deal with the worst-case scenarios. Due to its complexity, the tail of the \gls{paoi} distribution is a mostly uninvestigated subject in the literature, except for simple cases with 1 or 2 nodes~\cite{champati2019statistical}. In this section, we give bounds for the \gls{paoi} tail in the $K$-node scenario with intermediate traffic in the error-free case.
We know that the \gls{paoi} $\xi_i$ is given by $T_i+Y_i$. Since $T_i=W_i+S_i$, we use the upper bound on $W_i$ from~\eqref{eq:upper}:
\begin{equation}
 \xi_i\leq T_{i-1}+S_i+Y_i.
\end{equation}
Since $T_{i-1}$, $Y_i$, and $S_i$ are independent sums of exponential variables, the bound on $\xi_i$ is a hypoexponential random variable with a vector $\bm{\omega}$ of parameters $(\bm{\alpha},\bm{\mu},\lambda)$. We can sort the parameters in vector $\bm{\omega}$ of length $M$, in which each element $\omega_i$ is unique. The number of appearances of $\omega_i$ in the original vector is denoted as $m_i$. The \gls{cdf} of this hypoexponential random variable is given by:
\begin{equation}
 F(\tau)=\sum_{\ell=1}^M\sum_{n=1}^{m_\ell}\nu_{\ell,n}\left(\frac{1}{\omega_\ell^n}-\sum_{j=0}^{n-1}\frac{\tau^j e^{-\omega_\ell\tau}}{j!\ \omega_\ell^{n-j}}\right),
\end{equation}
where $\nu_{\ell,n}$ represents the hypoexponential coefficient as computed in~\eqref{eq:hypo_coeff} using vector $\bm{\omega}$.

\section{Numerical evaluation} \label{sec:results}
\begin{figure}[t!]
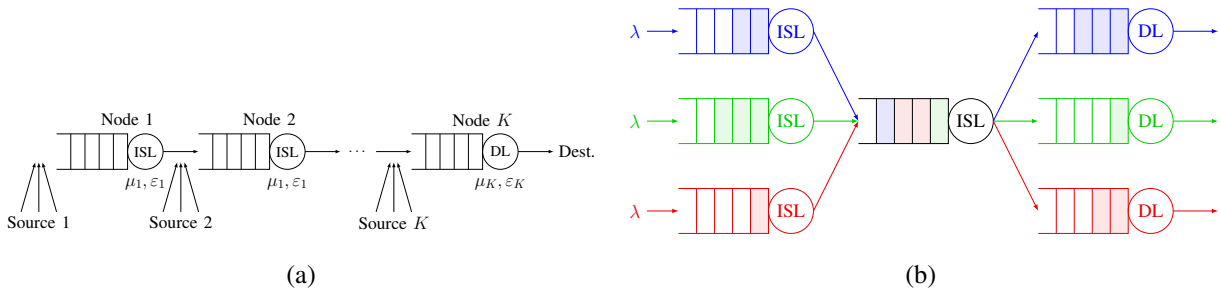

	\centering
	\begin{subfigure}[b]{0.49\linewidth}
	\resizebox{\linewidth}{!}{
        \includegraphics{./line.tex}}	
        \caption{}
        \label{fig:ln}
    \end{subfigure}
	\begin{subfigure}[b]{0.49\linewidth}
	\resizebox{\linewidth}{!}{
        \includegraphics{./dumbbell.tex}}
        \caption{}
        \label{fig:db}
	\end{subfigure}\vspace{0.5cm}
 \caption{Instances of the general model in Fig. \ref{fig_system_model} for the numerical evaluations: (a) Line network with a single destination and ground traffic at each node. (b) Dumbbell network with three source-destination pairs sharing a link.}
 \label{fig_linedumbbell}
\end{figure}

In order to verify the correctness of the theoretical results, we simulated the two scenarios in Fig.~\ref{fig_linedumbbell} using a Monte Carlo approach. The scenarios are instances of the general one depicted in Fig.~\ref{fig_system_model}, and they represent two different possible configurations in a \gls{leo} satellite network.
\begin{itemize}
    \item The \emph{line network} represents a scenario in which ground nodes placed in a remote area report to a ground station through a chain of $K$ satellites. It corresponds to Fig.~\ref{fig_system_model}a: at each satellite in the connection, an aggregated source with packet generation rate $\theta_i$ sends packets to the same destination, i.e., $\psi_i=0\,\forall i\leq K$. The bottleneck is the downlink between the last satellite and the ground station receiver, as it needs to serve traffic from all sources.
    \item The \emph{dumbbell topology} represents a scenario in which multiple connections share a single \gls{isl}, and then have different destinations. In this case, the shared \gls{isl} represents the shared bottleneck. If the $k$-th \gls{isl} is the shared one, we have $\theta_j=0$ $\forall j\neq k$ and $\psi_k=1$. This scenario is represented in Fig.~\ref{fig_system_model}b, and it can happen e.g. for inter-plane links of a constellation, as traffic between the two orbital planes is concentrated in the best link between them.
\end{itemize}
These two scenarios represent two extreme situations: in the former, cross traffic accumulates all the way to the final link, while in the other, it is concentrated in a single \gls{isl}, while all other links are less loaded. By mixing the two scenarios and considering partially overlapping paths, we can represent any realistic network with multiple sources and destinations. The full parameter list we used is in Table~\ref{tab:paraml}.

\begin{table}[b]
 \centering
 	\caption{Simulation parameters.\vspace{0.25cm}}
	\begin{tabular}{ccl}
		\toprule
		Parameter & Value &	Description \\
		\midrule
		$K_{\text{line}}$ & $\{2,6,10\}$ & Number of satellite relays for the line network\\
		$\mu_{\text{ISL}}$ & 1 & Service rate of the \gls{isl} links \\
		$\mu_{\text{DL}}$ & 0.8 & Service rate of the downlink\\
		$\psi$ & 0 & cross traffic leaving the connection after each node\\
		$\varepsilon$ & 0.01 & Error probability for all links\\
		$N_{\text{pkt}}$ & 100000 & Total number of packets for each source\\
		$K_{\text{db}}$ & $4$ & Number of satellite relays for the dumbbell topology\\
		$N_{\text{db}}$ & $\{2,6,10\}$ & Number of cross traffic sources for the dumbbell topology\\
		\bottomrule
	\end{tabular}
	\label{tab:paraml}
\end{table}

\subsection{Line network}

In the line network scenario, depicted in Fig.~\ref{fig:ln}, we simulated a network with a variable number of satellite relays and ground source traffic at each node. 
In the figure, multiple sources are present for all \gls{leo} satellites, but we consider the aggregated traffic per node $\lambda_k$ in the following. We analyzed both the error-free and error-prone cases, setting a constant error probability for all links. In each simulation, we discarded the initial transition to the steady state and the final packets to ensure that the results reflected the steady state behavior of the system. 

We considered results as a function of the error-free load $\rho$, which in our case is given by:
\begin{align}
    \rho&=\frac{\lambda+\sum_{j=1}^K\theta_j}{\mu_{\text{DL}}}.
\end{align}
In our simulation, we assume $\theta_1=0$, and that $\lambda=\theta_j=\rho\mu_{\text{DL}}/K$, $\forall j\in\{2,\ldots,K\}$. We do not consider the error in our computation of $\rho$, in order to provide a meaningful comparison between the error-free and error-prone cases. 

\begin{figure}[t!]
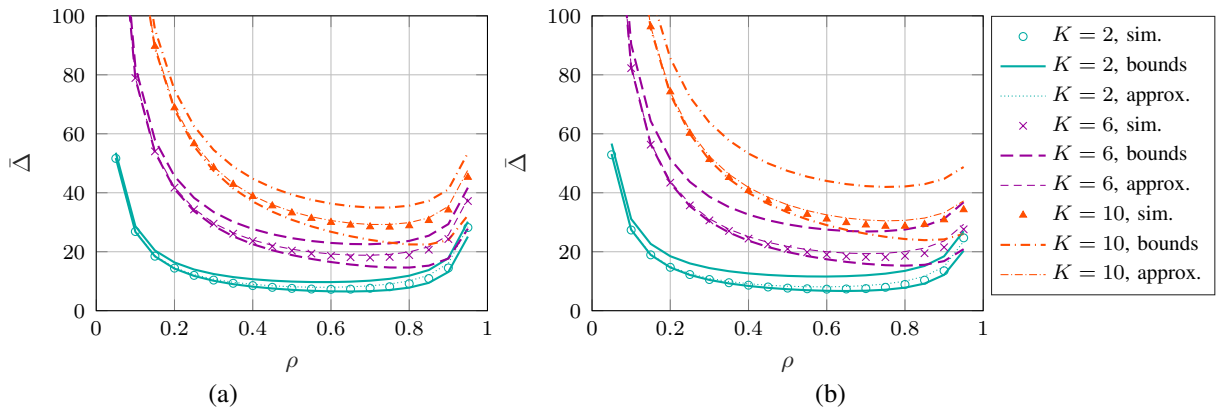

	\centering
	\begin{subfigure}[l]{0.36\linewidth}
        \includegraphics{./fcfs_noerror.tex}\vspace{-0.4cm}
        \caption{}
        \label{fig:fcfs_line_noerr}
	\end{subfigure}\hspace{0.5cm}
	\begin{subfigure}[l]{0.54\linewidth}
        \includegraphics{./fcfs_error.tex}\vspace{-0.4cm}
        \caption{}
        \label{fig:fcfs_line_err}
	\end{subfigure}\hspace{0.75cm}\vspace{0.5cm}
 \caption{Average \gls{aoi} as a function of the maximum load for the \gls{fcfs} policy in an error-free (left) and error-prone (right) line network.}
 \label{fig:fcfs_line}
\end{figure}

We first evaluate the average \gls{aoi} in the error-free scenario and the \gls{fcfs} policy, computing the upper and lower bounds. Unlike the system delay, the \gls{aoi} follows a U-shaped curve, as Fig.~\ref{fig:fcfs_line_noerr} shows. If the traffic load is very low, the average \gls{aoi} is very high, as the dominant factor is the time between successive packets from the same source. For instance, when $\rho=0.05$ and $K=10$, the arrival rate $\lambda_k$ for each source is just 0.004, as $\mu_{\text{DL}}=0.8$. This means that the average interarrival time is 250. As $\rho$ increases, the interarrival time decreases, but queueing becomes the main source of delay for very high traffic loads. The bounds we derived for the average \gls{aoi} are very tight for low values of $\rho$, as the approximation on the queueing time has a limited effect on the overall \gls{aoi}. If $\rho$ is high, the bounds are still reasonably tight, particularly for low values of $K$. The approximation based on the independence hypothesis is instead very close to the empirical curve even for high values of $\rho$, with a slight overestimation of the average \gls{aoi}.

Fig.~\ref{fig:fcfs_line_err} shows the average \gls{aoi} for the \gls{fcfs} policy in the error-prone case. While the overall results are similar, there are some differences between the two cases: errors increase the \gls{aoi} when $\rho$ is low, as the loss of one of the already rare packets can cause a significant increase. However, errors can actually have a beneficial impact on the average \gls{aoi} in high traffic load scenarios: since packets are frequent, one loss does not significantly increase the \gls{aoi}, and the reduced load on the downlink can improve the congestion and decrease queueing delays. In this case, the bounds are still tight for low values of $\rho$, but the upper bound is looser for high values of $\rho$. The approximation still very slightly overestimates the average \gls{aoi}, but it fits well the trend, particularly for low values of $\rho$. As we stated earlier, these simulations use an infinite buffer, but the impact of using a limited buffer is negligible, below 1\% of the average \gls{aoi} in all cases, except when using a buffer of just 1 or 2 packets.

\begin{figure}[!t]
 \centering
 \includegraphics{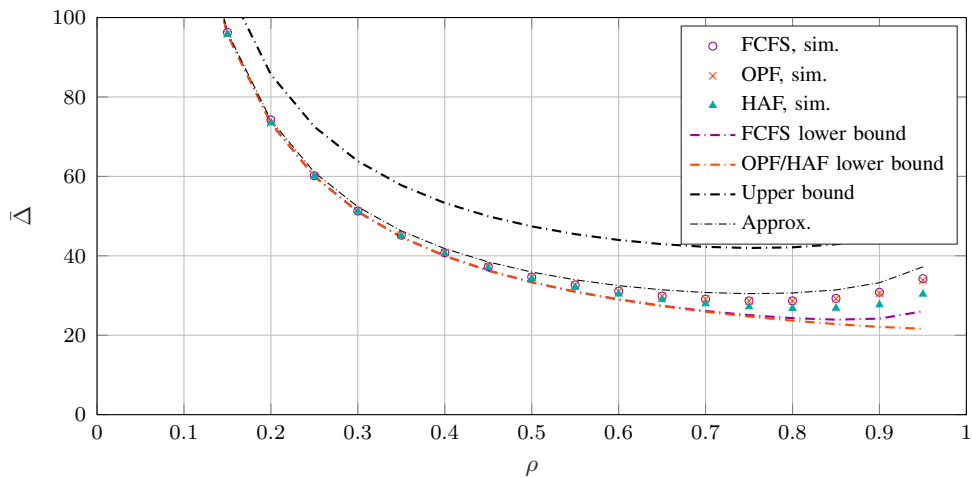}
 \caption{Average \gls{aoi} as a function of the maximum load for the \gls{fcfs}, \gls{opf}, and \gls{haf}
 policies in a line network with errors, $K=10$.}
 \label{fig:opf_err}
\end{figure}

\begin{figure}[!t]
 \centering
 \includegraphics{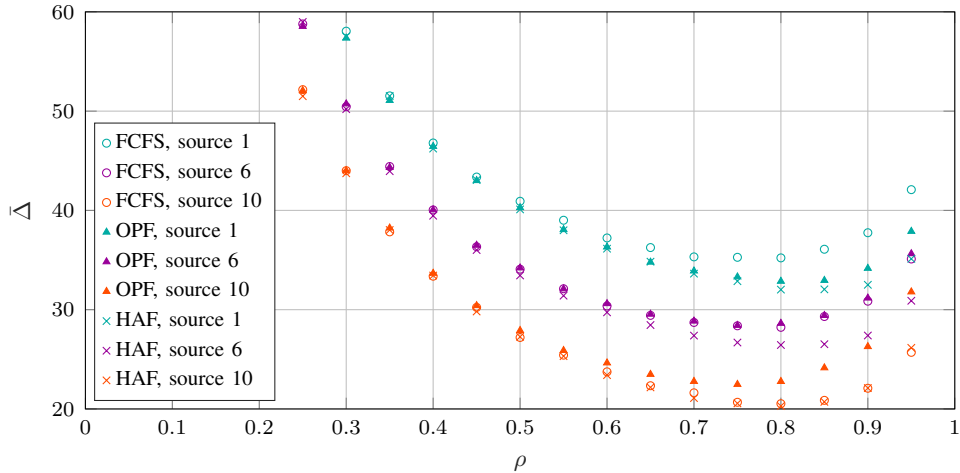}
 \caption{Average AoI for the first, sixth and last source for the \gls{fcfs}, \gls{opf}, and \gls{haf} policies in a line network with errors, $K=10$.}
 \label{fig:age_by_source}
\end{figure}

We then consider the impact of the scheduling by evaluating the \gls{opf} and \gls{haf} policies, whose \gls{aoi} performance is shown in Fig.~\ref{fig:opf_err}. The difference between \gls{opf} and \gls{fcfs} in terms of average \gls{aoi} for all sources is negligible, while \gls{haf} manages to slightly reduce the \gls{aoi} if the traffic load is high. As discussed in the previous section, the lower bound for the \gls{opf} and \gls{haf} policies is the same, and the upper bound is the same for all policies. The lower bound for \gls{fcfs} is slightly tighter, as it relies on fewer simplifying assumptions. 

\begin{figure}[!t]
 \centering
 \includegraphics{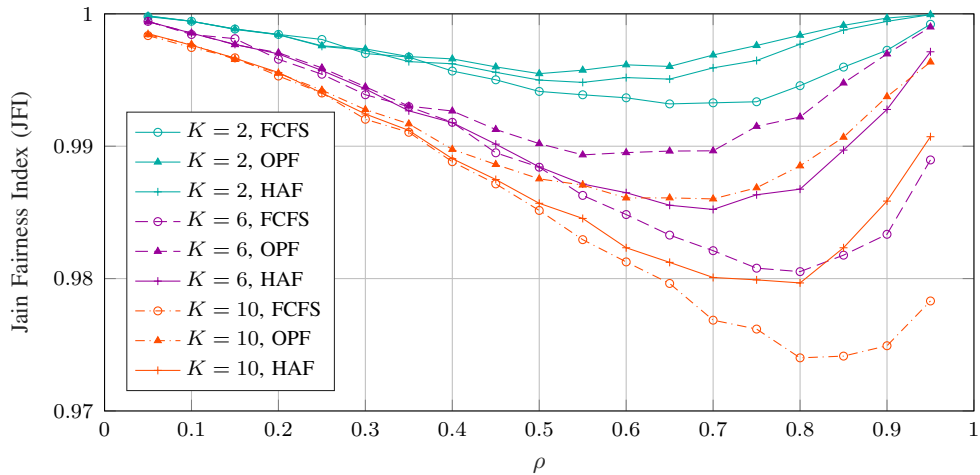}
 \caption{JFI as a function of the maximum load for the \gls{fcfs}, \gls{opf}, and \gls{haf} policies in a line network with errors.}
 \label{fig:jfi}
\end{figure}

The difference between the \gls{opf} and \gls{fcfs} policies is mostly based on fairness: while \gls{fcfs} nodes do not consider the delay that packets have accumulated on previous links, \gls{opf} bases its decisions on packet timestamps, reducing the \gls{aoi} distance between the ground sources close to the destination and the ones at the beginning of the chain. On the other hand, \gls{haf} is more efficient at preventing a surge of packets from few sources from increasing the \gls{aoi} for all others, as it prioritizes sources with the highest measured \gls{aoi} at each node. 
These fairness observations are confirmed in Fig.~\ref{fig:age_by_source} and Fig.~\ref{fig:jfi}. Fig.~\ref{fig:age_by_source} shows the average \gls{aoi} as a function of $\rho$ for three different sources: at the beginning, at the middle and at the end of the chain. For all policies, the first source is the one with the highest \gls{aoi}, as it has to traverse more links, but privileging older packets reduces this effect, allowing the packets from sources farther away from the destination to jump to the front of the line if they have already suffered significant delays. \gls{haf} can achieve the same \gls{aoi} as \gls{opf} for the first source, without the \gls{aoi} increase for later ones: it is consistently better than the other two policies, as we also remarked when analyzing the average \gls{aoi} across the whole network. 
Fig.~\ref{fig:jfi}, which shows the \gls{jfi} for the three policies in the considered scenario: the difference between sources is minimal when the traffic load is low, but increases as queueing becomes a factor. \gls{opf} can significantly reduce this difference, and its effects are starker for longer satellite relay chains. The \gls{haf} policy has an intermediate \gls{jfi}, as it can increase fairness with respect to the \gls{fcfs} policy, but not as much as \gls{opf}. 

\begin{figure}[!t]
 \centering
 \includegraphics{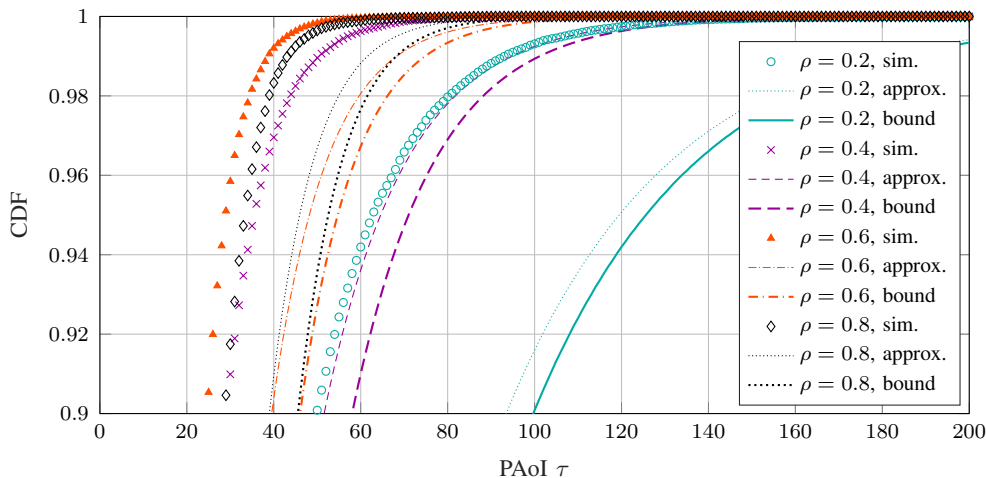}
 \caption{Empirical \gls{cdf} and upper bound on the tail of the distribution of the \gls{paoi} of the first source for a line network with $K=6$ in the error-free scenario (FCFS policy).}
 \label{fig:tail_k6}
\end{figure}

We then look at the bound on the tail of the distribution in the error-free scenario. We compare the empirical \gls{cdf} of the \gls{paoi} in the Monte Carlo simulations with the upper bound we computed in Sec.~\ref{sec:tail}. Fig.~\ref{fig:tail_k6} shows the empirical \glspl{cdf} and the theoretical bounds for $K=2$ and different values of $\rho$. In this case, the bound is almost always loose, except for $\rho=0.8$, and even the approximation acts as an only slightly tighter bound. We only show the \gls{fcfs} policy, but the others have a similar tail distribution. The looseness of the bounds can be explained by the fact that they are derived by decoupling $Y_i$ and $W_i$: while this is not an unrealistic assumption in the average calculation, the negative correlation between the two is very important in the tail of the distribution, as combinations of long interarrival times and long waiting times are extremely rare in practice, but not in the upper bound distribution.

\begin{figure}[!t]
 \centering
 \includegraphics{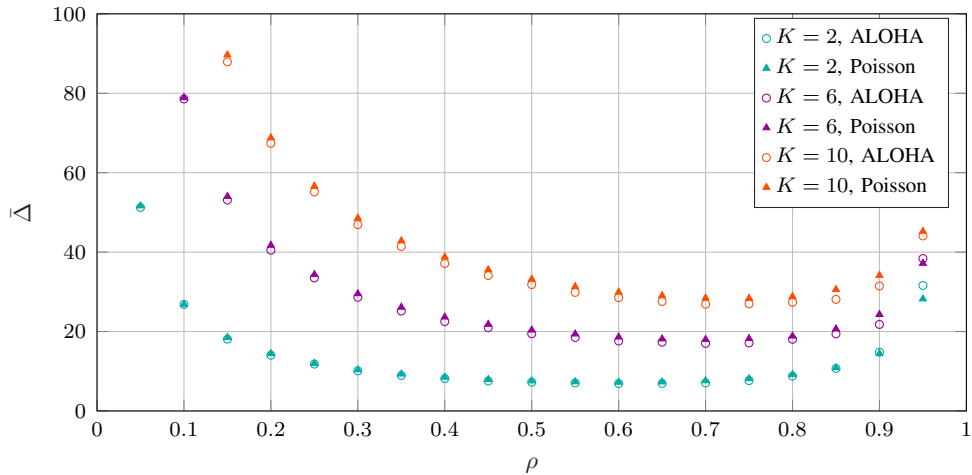}
 \caption{Comparison between the Monte Carlo measured \gls{aoi} for Poisson arrival model and the output of a realistic ALOHA uplink.}
 \label{fig:aloha}
\end{figure}

Finally, Fig.~\ref{fig:aloha} compares the average \gls{aoi} with ideal Poisson arrivals and the one that results from a realistic ALOHA uplink: if the rate at the first \gls{isl} is the same, the ALOHA uplink gets a slightly lower \gls{aoi} than Poisson arrivals, except for very high values of the load. This might be due to the second-order statistics of the arrival distribution, but it warrants more future analysis. We remark that having the same rate at the satellite means that the sources' actual packet generation rates are much higher, as the ALOHA uplink loses most of the transmitted packets because of collisions. In any case, the Poisson assumption allows us to draw accurate conclusions about the behavior of the system.

\subsection{Dumbbell topology}

In the dumbbell topology scenario, we consider $K=4$, with cross traffic on the second \gls{isl}, i.e., $\theta_2>0$ and $\psi_2=1$. We consider a number of sources $N$, each with packet generation rate $\lambda$, so that the total error-free load on the bottleneck is $\rho=N\lambda$.

\begin{figure}[!t]
 \centering
 \includegraphics{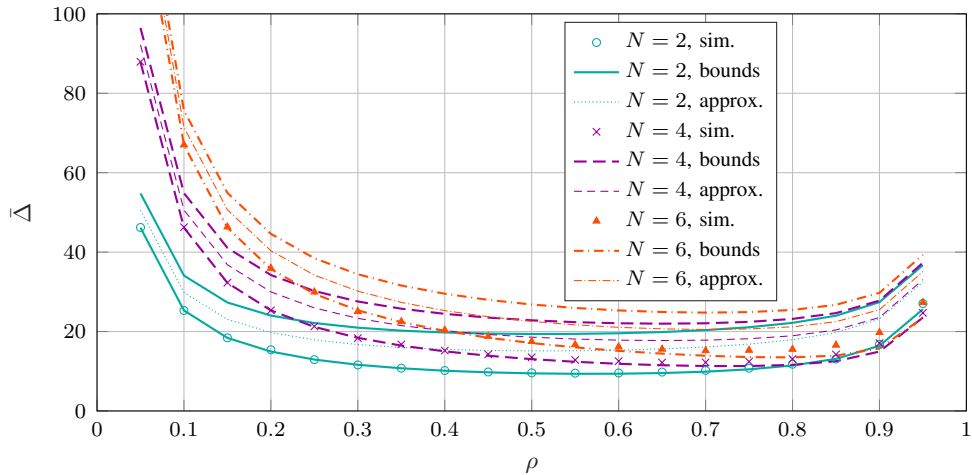}
 \caption{Average \gls{aoi} as a function of the maximum load for the \gls{fcfs}
 policy with 2, 4, and 6 sources in a dumbbell network with errors, $K=4$.}
 \label{fig:db_fcfs}
\end{figure}

As for the line network topology, we first examine the average \gls{aoi} when using the \gls{fcfs} policy. We set $\varepsilon_j=0.01\,\forall j$, as in the line network. The average \gls{aoi} (which is the same for all sources, as the network is symmetrical) is shown in Fig.~\ref{fig:db_fcfs}. As for the line network, the approximation appears to overestimate the \gls{aoi}, while the lower bound is a tight fit.

Interestingly, networks with a larger number of sources have their minimum \gls{aoi} with a higher load, as the effect of interarrival times is stronger when the same traffic $\rho$ is generated by multiple sources: if we set $\rho=0.7$, $\lambda=0.28$ when $N=2$, but $\lambda=0.093$ for $N=6$. Another interesting pattern shows that the \gls{aoi} increases for any number of sources for very high values of $\rho$, but it is less pronounced for a larger $N$. Our analyses led us to the conclusion that a lower number of sources, and a consequently higher $\lambda$ for each one, can lead to queueing delay on the first link, while a slightly lower value of $\lambda$ can ensure a smoother path until packets reach the bottleneck. This leads us to an important consideration when routing in \gls{leo} networks: the bottleneck should be placed as early as possible in the path, as links before might suffer from queueing, but packets coming out from the bottleneck are spaced far apart in time and are almost never queued at later links. The line network example we presented above, with gradually increasing load until the bottleneck in the downlink, is the worst possible scenario for \gls{aoi}.

\begin{figure}[!t]
 \centering
 \includegraphics{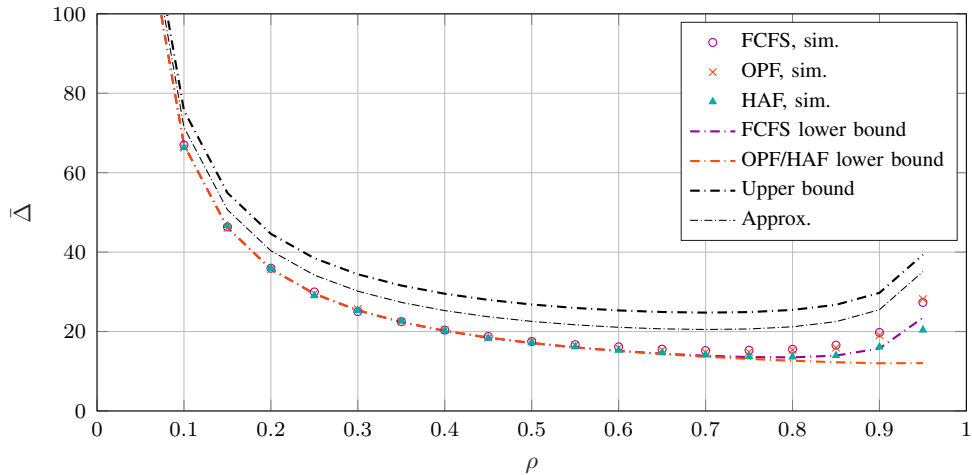}
 \caption{Average \gls{aoi} as a function of the maximum load for the \gls{fcfs}, \gls{opf}, and \gls{haf}
 policies with 6 sources in a dumbbell network with errors, $K=4$.}
 \label{fig:db_policy}
\end{figure}

\begin{figure}[!t]
 \centering
 \includegraphics{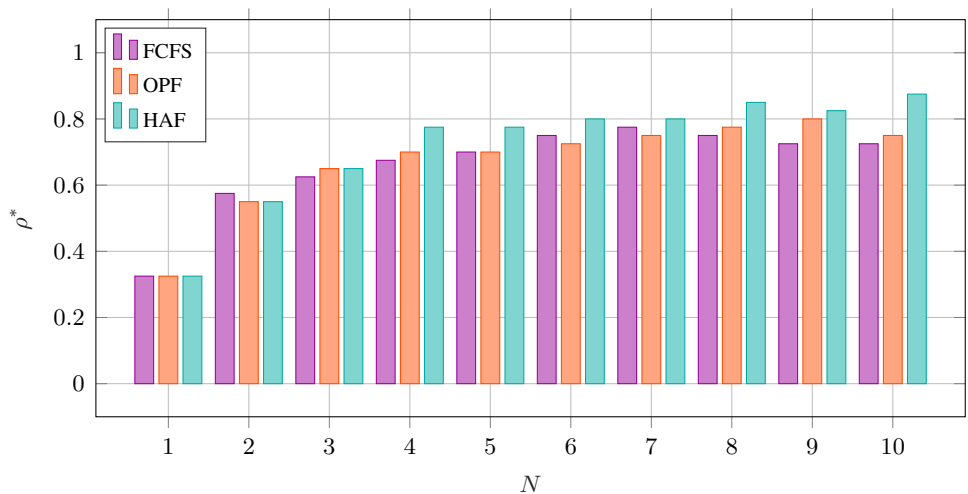}
 \caption{Optimal $\rho^*$ to minimize \gls{aoi} as a function of the number of sources for the three policies in a dumbbell network with errors, $K=4$.}
 \label{fig:db_opt}
\end{figure}

Fig.~\ref{fig:db_policy} shows the effect of applying different policies for $N=6$: as in the line network, \gls{opf} has no significant effect on the average \gls{aoi}, while \gls{haf} can perceptibly reduce the average \gls{aoi}, particularly for higher loads. In the dumbbell topology, the symmetry of the scenario makes fairness considerations moot: since all sources see the same cross traffic and connection parameters, the fairness is perfect. Interestingly, the \gls{haf} policy also works better if the load is higher, as Fig.~\ref{fig:db_opt} shows: for large numbers of sources, the traffic for each source is relatively low, and the critical task is coordinating among sources. If the rate is higher, there are almost always packets from all sources, with lower interarrival times, and the \gls{haf} policy can schedule sources fairly. On the other hand, \gls{fcfs} and \gls{opf} are more vulnerable to surges of packets from a subset of sources, as they have no way to give priority to sources with a high \gls{aoi}.

\section{Conclusions and future work} \label{sec:conclusions}

In this work, we modeled a network of \gls{leo} satellite relays as a tandem queue with $K$ nodes, and derived analytical bounds on the average and tail \gls{aoi}. The study of this kind of systems has been very limited, as most of the research on \gls{aoi} concentrates on single- or 2-hop connections, and the complexity of the scenario can make the derivations extremely unwieldy. The bounds we found are relatively tight, and lay the groundwork for more precise formulations. 

There are several possible avenues of future research on the subject, which include congestion control schemes limiting the sources' packet generation rates to maintain the lowest \gls{aoi} for the system, as well as considering queue management schemes such as preemption in scenarios with mixed traffic of short status updates and long transmissions, which has been studied extensively in single-source systems but is still largely unexplored for the multi-source case. Another possibility is to consider a more realistic model for the \glspl{isl}, which complicates the analysis significantly by making each relay a $G/G/1$ system. Finally, the derivation of tighter bounds on the tail of the distribution, which might include the error-prone scenario, should be a priority for researchers, as worst-case design is a critical element of future reliable applications.



\bibliographystyle{IEEEtran}
\bibliography{JSAC_AoI_relay}

\end{document}